# Complex Lasers with Controllable Coherence

Hui Cao[1]*, Ronen Chriki[2], Stefan Bittner[1], Asher A. Friesem[2], Nir Davidson[2]*

[1] Department of Applied Physics, Yale University, New Haven, CT, 06520, USA

[2] Department of Physics of Complex Systems, Weizmann Institute of Science, Rehovot 7610001, Israel

* Corresponding authors: hui.cao@yale.edu, nir.davidson@weizmann.ac.il

**Abstract**

The invention of lasers 60 years ago is one of the greatest breakthroughs in modern optics. Throughout the years, lasers have enabled major scientific and technological advancements, and have been exploited in numerous applications due to their advantages such as high brightness and high coherence. However, the high spatial coherence of laser illumination is not always desirable, as it can cause adverse artifacts such as speckle noise. To reduce the spatial coherence, novel laser cavity geometries and alternative feedback mechanisms have been developed. By tailoring the spatial and spectral properties of cavity resonances, the number of lasing modes, the emission profiles and the coherence properties can be controlled. This review presents an overview of such unconventional, complex lasers, with a focus on their spatial coherence properties. Laser coherence control not only provides an efficient means for eliminating coherent artifacts, but also enables new applications.

# 1. Introduction

## 1.1. Conventional lasers

Lasers have been widely used in industry, medicine and other areas of contemporary life due to their high brightness, high coherence, high efficiency and good spectral control. The essential elements of a laser are a material that amplifies light through stimulated emission (the gain medium) and a cavity that traps light in the gain medium to enable more efficient amplification[1]. The gain medium is pumped electrically by a current source or optically by a flash lamp or another laser source.

The classical laser cavity is composed of two spherical mirrors facing each other[1]. A gain medium is placed in between the mirrors such that light bouncing back and forth between the mirrors is amplified during each passage through the gain medium. At least one of the spherical mirrors is only partially reflective, so a portion of the light leaves the cavity every round trip, thereby producing an output beam. When the light amplification (gain) is strong enough to compensate for loss due to cavity leakage and material absorption, lasing oscillations occur in the cavity.

After propagating one round trip through the cavity, the spatial field profile must be identical to the original one. This resonance condition determines the field distributions of cavity modes. The number of field nodes in the longitudinal direction (along the cavity axis) gives the longitudinal quantization number, also called the longitudinal mode number. Two transverse mode numbers give the numbers of field nodes in the two orthogonal transverse directions (perpendicular to the cavity axis). Modes with low (high) quantization numbers are referred to as low (high) order modes.

The light leakage from the cavity is characterized by the quality ($Q$) factor, which is proportional to the photon lifetime in the cavity and inversely proportional to the round-trip loss. If the cavity length (equal to the distance between the two spherical mirrors) is much larger than the cavity width (determined by the transverse dimension of the mirrors), light experiences diffraction loss while propagating from one mirror to the other. The higher order transverse modes have finer spatial features and larger transverse wave vector components, thus suffering stronger diffraction loss and having lower $Q$ factors than the lower-order modes.

If the gain material is uniformly distributed in the cavity, the lower-order transverse modes have lower thresholds and lase first at lower pump power. They can saturate the optical gain, especially for a homogeneously broadened gain medium, and thus prevent the higher-order modes from lasing at higher pump powers. This effect, known as mode competition, usually determines the number of transverse modes that lase simultaneously, thus having a profound impact on the spatial coherence of laser emission.

## 1.2. Laser coherence properties

The first-order coherence function describes correlations of the electric field $E$ in space and time[2], and is defined as

$$g^{(1)}(r_1, r_2, \tau) = \frac{|\langle E^*(r_1, t) E(r_2, t+\tau) \rangle|}{\langle E^*(r_1, t) E(r_1, t) \rangle^{1/2} \langle E^*(r_2, t+\tau) E(r_2, t+\tau) \rangle^{1/2}}$$

where $\langle ... \rangle$ denotes averaging over time $t$ for a duration longer than all intrinsic dynamical time scales. Specifically, the temporal coherence $g^{(1)}(\tau)$ describes the correlation of fields at the same location ($r_1 = r_2$) but different time, while the spatial coherence $g^{(1)}(r_1, r_2)$ describes the correlation of fields at the same time ($\tau = 0$) but different locations. The width of $g^{(1)}(\tau)$ gives the coherence time. It is inversely proportional to the spectral width of the laser emission, which depends on the number of longitudinal lasing modes and their linewidth. The width of $g^{(1)}(r_1, r_2)$ gives the coherence length, which is inversely proportional to the number of transverse lasing modes. Hence, the longitudinal modes are also called temporal modes, while the transverse modes are called spatial modes.

Commonly used lasers, such as edge-emitting diode lasers, fiber lasers and solid-state lasers, have only a few transverse lasing modes due to gain competition, thus their spatial coherence is usually high. It should be noted that the spatial coherence is independent of the number of longitudinal lasing modes with identical transverse structure. One way of reducing the spatial coherence is to increase the number of transverse lasing modes. This can be realized by reducing the ratio of cavity length to mirror diameter so that the diffraction loss for higher-order transverse modes is reduced, such as for vertical cavity surface emitting lasers (VCSELs). Other approaches will be discussed in section 2.

High spatial and temporal coherences are not exclusive to lasers. Spatial or spectral filtering of spontaneous emission from a lamp can greatly improve spatial or temporal coherence, albeit at the price of greatly reduced power. A fundamental difference between a laser and a lamp lies in the quantum statistical properties of their emission[3]. Specifically, on the time scale shorter than the characteristic time given by the inverse of the spectral bandwidth, laser light has Poissonian photon statistics, while thermal light has Bose-Einstein statistics.

### 1.3. Coherent artifacts

A laser with high spatial coherence can generate a directional light beam with small divergence, and the beam can be focused to a diffraction-limited spot. Unfortunately, high spatial coherence also causes deleterious effects such as coherent artifacts and cross-talk in full-field imaging and displays[4]. The most common manifestation of coherent artifacts is speckle noise[5,6]. A rough object or scattering environment introduces random phase delays among mutually coherent photons that interfere to produce a random grainy pattern[6]. In addition to parallel imaging and display applications, the coherent artifacts pose serious problems for laser applications in material processing, photolithography, holography, and optical trapping of cold atoms and colloidal particles.

Speckles result from spatial coherence: light fields at different locations are mutually coherent. When these fields are scattered by a rough surface or in a disordered medium, they will be scattered to the same position via different paths. Because they are mutually coherent and have random phase delays, these fields interfere to form a random grainy pattern. For a fully developed speckle pattern, the intensity contrast $C = \sigma_I/\langle I \rangle$ is equal to 1, where $\sigma_I$ is the standard deviation of the intensity and $\langle I \rangle$ is the average intensity[5,6].

### 1.4. Speckle suppression

Because speckles originate from spatial coherence, the most effective way to suppress speckles is to resort to sources of low spatial coherence, such as lamps or light-emitting diodes (LEDs), which have been used for most full-field imaging and display applications. Unfortunately, these sources have lower power per mode, poorer collection efficiency, and lesser spectral control as compared to lasers. The power limitations are particularly problematic in imaging applications that involve absorbing or scattering media, prompting the use of raster-scanning modalities in laser-based imaging systems. However, scanning is time consuming and thus not suitable for imaging of moving objects or transient processes.

An alternative way to suppress speckles is to reduce the spatial coherence of light emitted from a laser, e.g., using a time-varying scattering system such as a rotating diffuser[7], a colloidal solution[8], or a micro-electromechanical mirror array[9]. In the case of a rotating diffuser, different angular positions of the diffuser produce uncorrelated speckle patterns. Summing the intensities of $N$ uncorrelated speckle patterns averages the speckle noise and reduces the speckle contrast[5,6] by a factor of $\sqrt{N}$ as illustrated in Fig. 1a. However, lowering the speckle contrast below the level of human perception[10,11] ($C < 3\%$) requires summing about 1000 different speckle patterns. Since it takes time to create that many speckle patterns, the imaging speed is limited by the velocity of the mechanical motion of the scattering system.

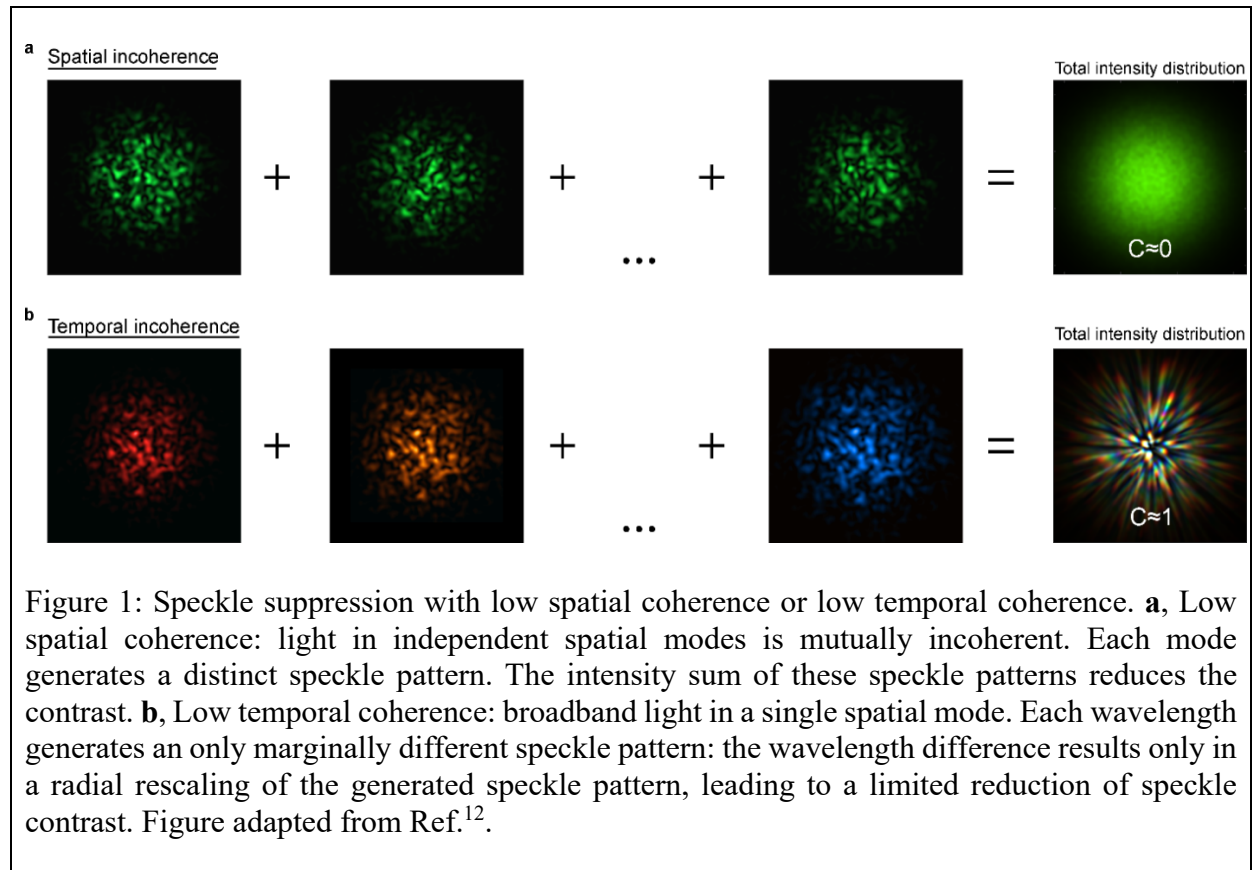

Figure 1: Speckle suppression with low spatial coherence or low temporal coherence. **a**, Low spatial coherence: light in independent spatial modes is mutually incoherent. Each mode generates a distinct speckle pattern. The intensity sum of these speckle patterns reduces the contrast. **b**, Low temporal coherence: broadband light in a single spatial mode. Each wavelength generates an only marginally different speckle pattern: the wavelength difference results only in a radial rescaling of the generated speckle pattern, leading to a limited reduction of speckle contrast. Figure adapted from Ref.[12].

Broadband sources with low *temporal* coherence, such as superluminescent diodes (SLDs) and supercontinuum sources, have also been exploited for speckle suppression. These sources have relatively high spatial coherence, and only moderate speckle suppression could be achieved[6]. Figure 1b is an illustration of broadband light illuminating a static diffuser. The speckle patterns exhibit a radial scaling with the wavelength. Thus summing the intensities of speckle patterns for different wavelengths has a limited effect on speckle suppression. Quantitatively, the spectral correlation function characterizes the degree of correlations among speckle patterns for different wavelengths. Only when the illumination bandwidth $\Delta\omega$ is much larger than the spectral correlation width $\delta\omega$, the spectral compounding is sufficient for speckle reduction. The speckle contrast $C \approx 1/\sqrt{\Delta\omega/\delta\omega}$ reaches 3% for $\Delta\omega \sim 1000 \cdot \delta\omega$. For a standard diffuser, the spectral correlation width is typically a few nanometers in wavelength[13], thus requiring an extremely broadband source. A possible solution is to drastically reduce $\delta\omega$ using multiple scattering in a thick diffusive medium or a long multimode fiber. However, it is difficult to adopt this approach for applications that rely on narrowband lasers such as excimer lasers for deep UV photolithography, or that require compact sources such as red-green-blue (RGB) lasers for portable displays.

Although various approaches have been developed over the years to suppress the deleterious coherence artifacts, they are mostly applied outside the lasers themselves. Such approaches can be summarized as starting with a laser of high spatial coherence, and then reducing the spatial coherence of its emission. A direct, more efficient approach is to resort to a laser with low spatial coherence, as will be discussed in section 2. Moreover, it is possible to control and tune the spatial coherence of a laser so as to enable new applications, as will be described in section 3.

## 2. Lasers with low or tunable spatial coherence

To overcome the natural tendency of high spatial coherence exhibited by conventional lasers and achieve low and/or tunable spatial coherence, fundamental changes in the laser design or operation are necessary to achieve low or even tunable spatial coherence. In this section we discuss three strategies to reduce the spatial coherence of lasers. Since only modes with different spatial emission profiles contribute significantly to the reduction of spatial coherence, the first strategy is to drastically increase the number of transverse lasing modes by modifying the laser cavity design. This is implemented, for example, in the degenerate cavity lasers presented in section 2.1 or the large mode-area fiber amplifier discussed in section 2.2.

The second strategy is a more radical approach to have many lasing modes with different spatial emission profiles. Most laser cavities, including those mentioned for the first strategy, feature a separable geometry with an optical axis that defines the dominant propagation direction, from which results the distinction into longitudinal and transverse modes. The second strategy instead relies on laser cavities without an optical axis or separable geometry so that individual lasing modes have no dominant propagation direction and can no longer be assigned longitudinal and transverse quantization numbers. Instead, each mode consists of many wave components with different directions, and as a result, all modes can have distinct emission profiles. This approach avoids having a number of different longitudinal lasing modes that do not, however, contribute to the reduction of spatial coherence. This second strategy has been implemented for example with cavities in the shape of chaotic billiards (see section 2.3) or with disordered scattering media (see section 2.4).

The main challenges for the aforementioned two strategies are to ensure that different spatial modes have similar $Q$ factors and that their competition for gain is minimal so they can actually lase simultaneously. An important advantage is that both strategies can be adapted to different gain materials.

The third strategy does not rely on exciting many distinct spatial modes, but instead aims at disrupting the formation of lasing modes in the first place. To this end, the cavity itself is not modified, but instead the pumping conditions are adjusted such that the cavity is constantly modified by thermal effects and the lasing modes cannot build up. Consequently, the emission is spatially incoherent (see section 2.5).

## 2.1. Degenerate cavity laser

For a large number of transverse modes to lase simultaneously, they must have similar lasing thresholds. This can be achieved with a degenerate cavity laser, in which all transverse modes have nearly identical $Q$ factors[14,15]. A degenerate cavity forms a self-imaging system, as exemplified in Fig. 2a. Two lenses in a 2f-2f telescope arrangement are placed between two flat mirrors so an arbitrary field distribution at any plane in the cavity is imaged onto itself after one round trip. Therefore, any field distribution is an eigenmode of the cavity, and all (orthogonal) eigenmodes are degenerate. Since the lasing thresholds of the transverse modes are nearly identical, all transverse modes can lase simultaneously. The number of transverse lasing modes is given by the ratio of the gain medium cross-section over the diffraction-limited area determined by the smallest numerical aperture (NA) of the optics inside the cavity. It has been shown that a degenerate solid-state laser can support more than $10^5$ transverse modes lasing simultaneously and independently, producing emission with extremely low spatial coherence[16].

To tune the degree of spatial coherence, the number of transverse lasing modes can be varied by inserting a variable circular aperture in the mutual focal plane of the two lenses, as shown in Fig. 2b. The aperture serves as a spatial filter that introduces loss to higher order transverse mode[16]. When the aperture is sufficiently small, only the lowest-order transverse mode lases. Since the spatial overlap of the lasing mode and the gain medium remains the same, the total power extracted from the gain medium is nearly unchanged. By decreasing the size of the aperture, the number of transverse lasing modes can be reduced from 320,000 to 1, while the output power decreases by less than 50%.

The remarkable efficiency of redistributing energy over $10^5$ transverse modes is unique for the degenerate cavity laser (Fig. 2c,d), and not possible with conventional laser resonators. Because the degenerate cavity modes can have arbitrary transverse profiles, there is no inherent transverse mode size, allowing the laser to adopt any mode size dictated by the aperture. For conventional stable laser resonators, however, the characteristic mode size is dictated by the cavity geometry, thus spatial filtering with, e.g., a circular aperture inevitably introduces loss in all modes due to the inherent mismatch. Traditionally, the spatial coherence of emission is tuned via spatial filtering outside the laser cavity, but then the total power reduces linearly with the number of transverse modes, assuming all modes have equal power. Hence, spatial filtering outside a laser is much less efficient than filtering inside a degenerate cavity.

A further manipulation of the spatial coherence properties can be obtained by resorting to more sophisticated intra-cavity spatial filters (masks) in the degenerate cavity laser[17]. For example, a variable slit enables independent control of the spatial coherence length in one coordinate axis without affecting that in the orthogonal axis; or two pinholes, an annular band or an array of circular holes can generate cosine, Bessel or comb-like spatial coherence functions, respectively. In principle, arbitrary spatial coherence functions can be obtained with a degenerate laser without losing power.

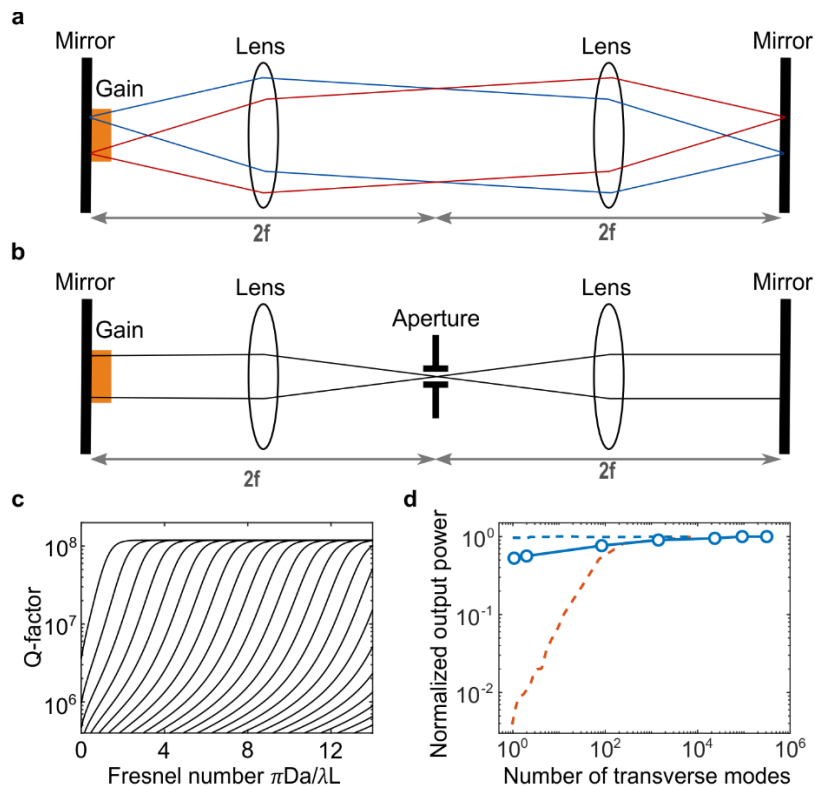

Figure 2: Degenerate cavity laser with tunable spatial coherence. **a**, Schematic of a degenerate laser cavity in a self-imaging configuration. Light emitted from any point in the gain medium will be imaged back to itself after one round trip. A large number of independent transverse modes lase simultaneously and independently to produce low spatial coherence. **b**, Inserting an aperture in the mutual focal plane of the two lenses yields a single transverse mode emitting with high spatial coherence. **c,** Cavity $Q$ factors as a function of the Fresnel number (proportional to the aperture diameter $a$) for an $L = 1$ m long degenerate cavity with an output coupler of 90% reflectivity and a gain medium with circular cross section of diameter 1 cm. As the aperture size increases, additional transverse modes appear and their $Q$ factors rise quickly to approach the value of existing modes. **d**, Measured output power as a function of the number of transverse modes in the degenerate cavity (solid blue line), compared to calculated output power vs number of modes for a degenerate cavity laser (blue dashed line) and a stable hemispherical cavity (red dashed line). The measurement was done with a degenerate cavity comprising two lenses with focal length of $f = 25$cm and an Nd:YAG crystal of 11 cm length and 0.95 cm diameter. Panels **a** and **b** adapted from Ref.[18]; panel **d** adapted from Ref.[16].

## 2.2. Multimode fiber amplifier

While it is easy to insert a solid-state or semiconductor gain medium into a degenerate cavity, it is hard to do so with a long optical fiber. Hence a different approach is needed for fiber lasers. A straightforward method for reducing the spatial coherence of a fiber laser is to increase the number of guided modes in the gain core, i.e., the fiber core doped with gain atoms. A recently developed multimode fiber with an extra-large mode area (XLMA) gain core supports numerous guided modes (Fig. 3a). However, the lower order modes are more tightly confined within the gain core, experiencing stronger amplification than the higher-order modes. Hence, the lower-order modes tend to deplete the gain and suppress the higher-order modes. Consequently, equalizing modal gain and reducing modal competition are necessary to obtain amplification of many guided modes. A simple way to equalize the modal gain is to introduce mode coupling. For example, fiber bending and twisting, or imperfections like fluctuations of the refractive index and variations of the fiber cross section cause random coupling among the modes and thus reduce mode-dependent gain.

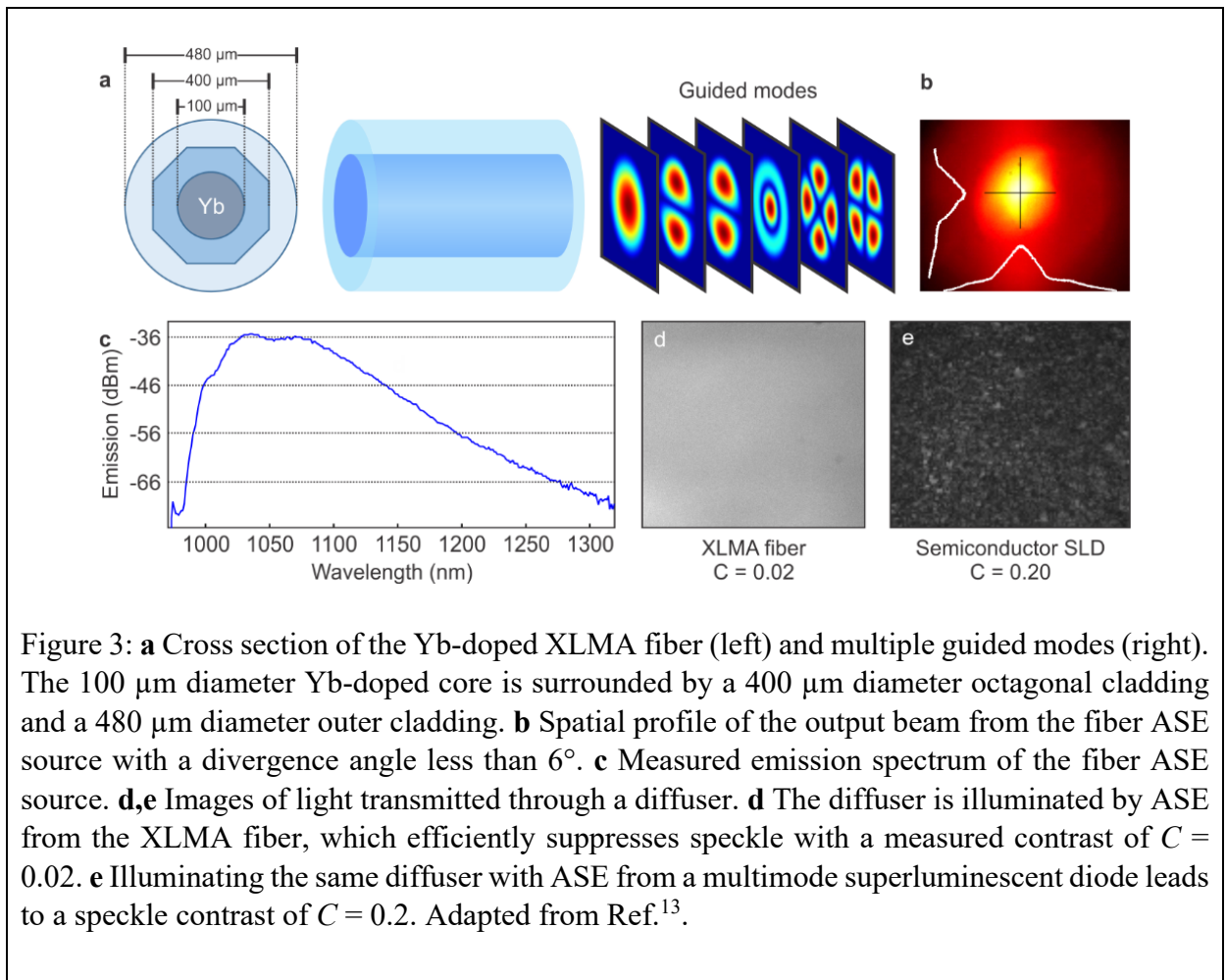

Figure 3: **a** Cross section of the Yb-doped XLMA fiber (left) and multiple guided modes (right). The 100 μm diameter Yb-doped core is surrounded by a 400 μm diameter octagonal cladding and a 480 μm diameter outer cladding. **b** Spatial profile of the output beam from the fiber ASE source with a divergence angle less than 6°. **c** Measured emission spectrum of the fiber ASE source. **d,e** Images of light transmitted through a diffuser. **d** The diffuser is illuminated by ASE from the XLMA fiber, which efficiently suppresses speckle with a measured contrast of $C = 0.02$. **e** Illuminating the same diffuser with ASE from a multimode superluminescent diode leads to a speckle contrast of $C = 0.2$. Adapted from Ref.[13].

One method for reducing modal competition is to operate in the regime of amplified spontaneous emission (ASE). This is achieved by cleaving the fiber end facets at an angle to minimize feedback for lasing[13]. A rare-earth-doped XLMA fiber ASE source has not only low spatial coherence (Fig. 3d), but also low temporal coherence due to its broad emission spectrum (Fig. 3c). In addition, the output beam is highly directional, with the divergence angle dictated by the small numerical aperture of the fiber. Despite the participation of many spatial modes, the spatial profile of the output beam is smooth, as shown in Fig. 3b. For comparison, semiconductor-based superluminescent diodes have relatively high spatial coherence (Fig. 3e), because they are based on one-dimensional waveguides that have much fewer transverse modes and much weaker mode mixing than the XLMA fiber[19].

2.3. Wave-chaotic microcavity lasers

Instead of increasing the number of transverse lasing modes, a more dramatic way of reducing the spatial coherence is breaking the separability into longitudinal and transverse modes so that all lasing modes have distinct emission profiles. In this way, all lasing modes contribute to the reduction of spatial coherence unless they are phase locked. To this end, a fundamental change of the laser cavity geometry is necessary.

Resonators with separable geometry such as the stable resonators discussed in section 1 or the circular cavity shown in Fig. 4a-c are integrable, that means the wave equation can be solved analytically and the resonant modes are labeled with sets of quantization numbers corresponding to the coordinate axes, e.g., radial and azimuthal indices in the case of the circle. Separable cavities also feature integrable classical ray dynamics: a small deviation of the initial conditions of two trajectories leads to an at most linear divergence.

Cavities with non-separable geometries, in contrast, can exhibit partially or completely chaotic ray dynamics[20,21], where small deviations of the initial conditions lead to an exponential divergence of trajectories. One example, shown in Fig. 4d, is a circular cavity with a section removed along a straight cut, leading to fully chaotic ray dynamics[22]. Since its shape resembles the letter D, it is called "D-cavity".

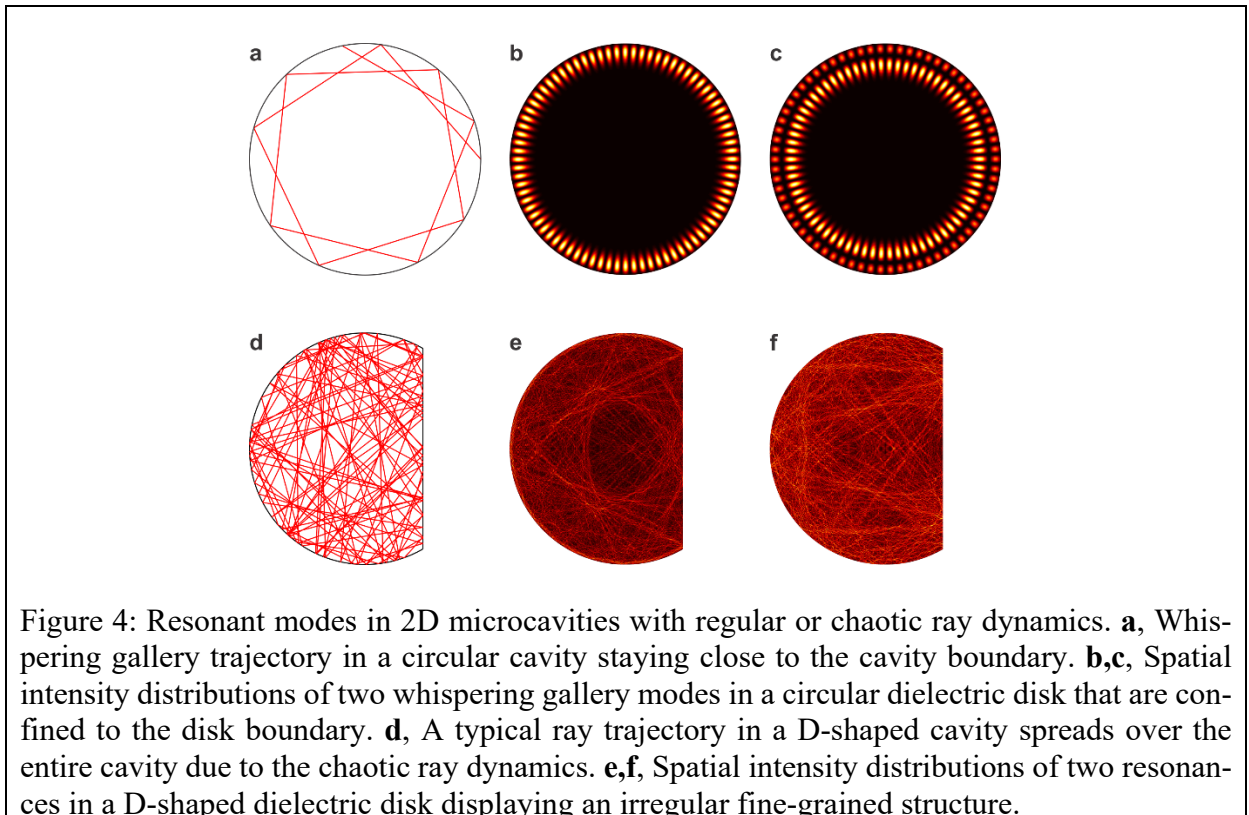

Figure 4: Resonant modes in 2D microcavities with regular or chaotic ray dynamics. **a**, Whispering gallery trajectory in a circular cavity staying close to the cavity boundary. **b,c**, Spatial intensity distributions of two whispering gallery modes in a circular dielectric disk that are confined to the disk boundary. **d**, A typical ray trajectory in a D-shaped cavity spreads over the entire cavity due to the chaotic ray dynamics. **e,f**, Spatial intensity distributions of two resonances in a D-shaped dielectric disk displaying an irregular fine-grained structure.

Regardless of integrable or chaotic ray dynamics, the passive cavity modes do not exhibit exponential dependence on the initial conditions, since the modes are solutions of the linear Helmholtz equation. However, the classical ray dynamics manifests in the mode properties according to the principle of ray-wave correspondence[23], and cavities with chaotic ray dynamics, called "wave-chaotic", show qualitative differences compared to integrable ones. First, the modes of wave-chaotic cavities cannot be labeled with sets of quantization numbers due to their non-separable geometry, and they are in general non-degenerate[20]. Moreover, their field distributions feature irregular patterns of wavelength-sized grains that can spread over the whole cavity just like the classical trajectories (see Fig. 4e,f), thus each mode has a distinct emission field profile. In contrast, modes of integrable cavities feature very regular field patterns (see Fig. 4b,c) that can have identical emission profiles if one of their quantization numbers matches. One important feature of the dielectric circle resonator are whispering gallery modes (WGMs) with extremely high $Q$ factors that correspond to trajectories near the circumference confined by total internal reflection (see Fig 4a).

The ray-wave correspondence also affects the lasing dynamics. Lasing in a circle is dominated by WGMs due to their very low thresholds. The number of lasing modes is limited by the strong gain competition of the WGMs caused by their localization near the boundary. In contrast, a D-shaped dielectric disk supports a large number of "chaotic" lasing modes with almost identical thresholds thanks to their relatively high but similar $Q$ factors[24]. Since their field distributions are typically widely spread over the cavity, their competition for gain is much weaker compared to WGMs. Consequently, many more modes can lase simultaneously in a D-cavity than in a circular cavity. All lasing modes have distinct emission profiles and thus contribute to the reduction of spatial coherence. By optimizing the position of the straight cut in the D-cavity, the modes spread more uniformly across the cavity to minimize their gain competition, and the differences of their lasing thresholds is further reduced. The maximum number of lasing modes in the D cavity is reached when the cut is half the radius away from the center[24].

Not all wave-chaotic cavities are necessarily well-suited for highly-multimode lasing. In addition to "chaotic" modes, they can exhibit scar modes that are localized on unstable periodic orbits[21,25]. If scar modes have higher $Q$ factors than the "chaotic" modes, they will dominate lasing, and their strong competition for gain (due to large spatial overlap) will limit the number of lasing modes. One advantage of the D-cavity is that all scar modes have low $Q$ factors because the angles of incidence of the underlying periodic orbits are below the critical angle for total internal reflection.

A disadvantage of the D-cavity laser for practical applications is that its emission is non-directional since its modes consist of wave components travelling in all possible directions just like the classical ray trajectories. Directional emission from a wave-chaotic cavity can be obtained by modifying the cavity geometry such that long-lived ray trajectories reach the critical angle of total internal reflection only at certain parts of the boundary and refract to particular directions[21,26]. Creating a wave-chaotic cavity that combines low spatial coherence with directional emission presents a future challenge.

2.4. Random lasers

In a wave-chaotic cavity, light is confined by reflection at the cavity boundary, so the cavity shape determines the spatial and spectral properties of the resonant modes. Alternatively light scattering induced by structural disorder can trap light and provide feedback for lasing even in the absence of a well-defined cavity[27-29]. As illustrated by Fig. 5a, in a gain medium that contains numerous scattering centers spontaneously emitted photons will be scattered many times and undergo a "random walk". Multiple scattering increases the path length of photons traveling in the gain medium, thus enhancing the stimulated emission of photons that amplifies light[30,31]. Furthermore, the scattered waves may return to spatial positions they have visited before, providing coherent feedback and enabling laser oscillation[32-34]. Such a laser is called a random laser. It has been implemented in various material systems, including powders[35], polycrystalline films[36], colloids[37], polymers[38], optical fibers[39] and organic materials[40]. The lasing frequencies range from ultraviolet[41] and visible[42] to infrared[43] and terahertz[44]. Although lasing is realized in both strong and weak scattering systems, the lasing threshold is lower when scattering is stronger[45].

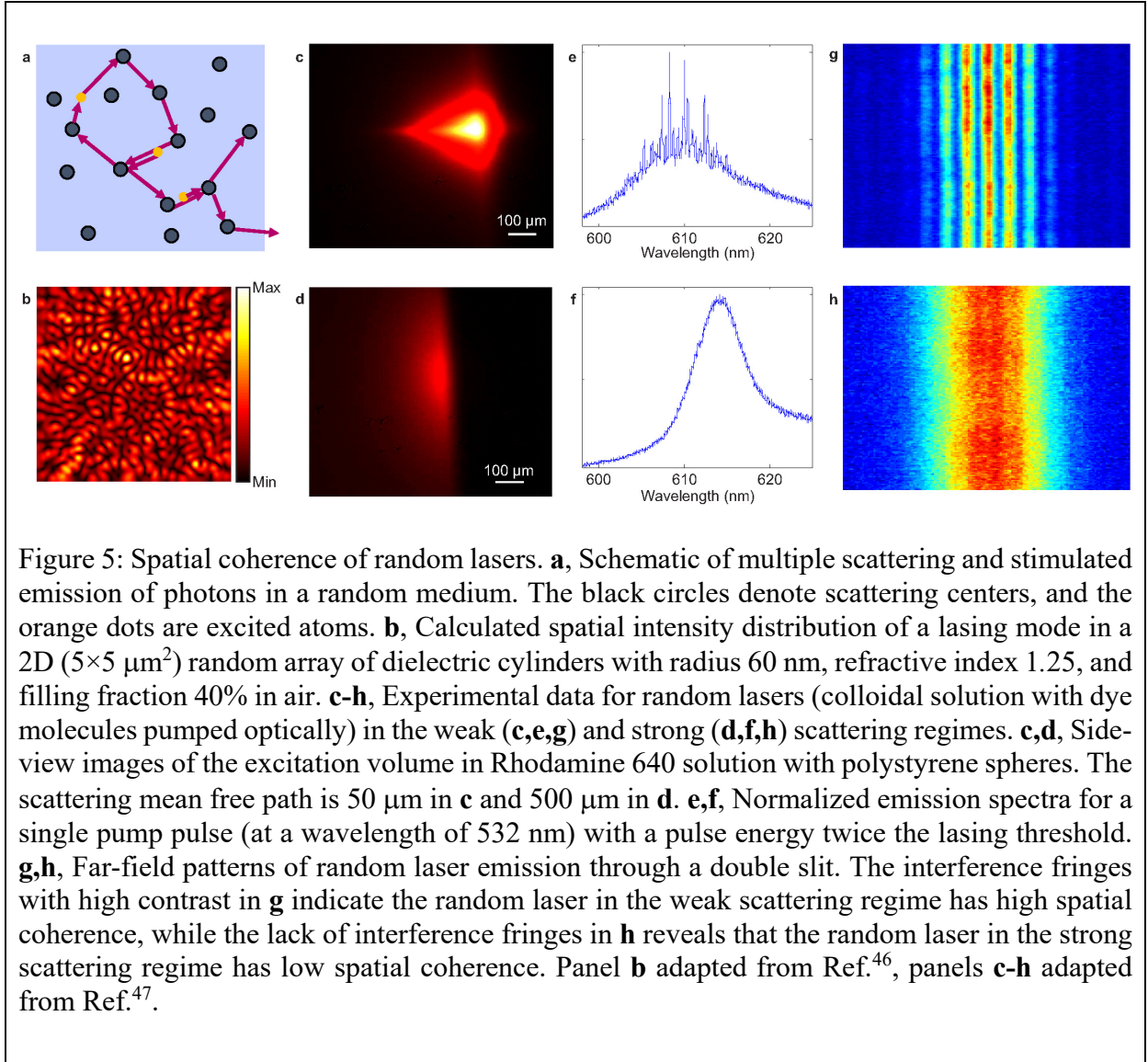

Figure 5: Spatial coherence of random lasers. **a**, Schematic of multiple scattering and stimulated emission of photons in a random medium. The black circles denote scattering centers, and the orange dots are excited atoms. **b**, Calculated spatial intensity distribution of a lasing mode in a 2D (5×5 $\mu m^2$) random array of dielectric cylinders with radius 60 nm, refractive index 1.25, and filling fraction 40% in air. **c-h**, Experimental data for random lasers (colloidal solution with dye molecules pumped optically) in the weak (**c,e,g**) and strong (**d,f,h**) scattering regimes. **c,d**, Side-view images of the excitation volume in Rhodamine 640 solution with polystyrene spheres. The scattering mean free path is 50 $\mu m$ in **c** and 500 $\mu m$ in **d**. **e,f**, Normalized emission spectra for a single pump pulse (at a wavelength of 532 nm) with a pulse energy twice the lasing threshold. **g,h**, Far-field patterns of random laser emission through a double slit. The interference fringes with high contrast in **g** indicate the random laser in the weak scattering regime has high spatial coherence, while the lack of interference fringes in **h** reveals that the random laser in the strong scattering regime has low spatial coherence. Panel **b** adapted from Ref.[46], panels **c-h** adapted from Ref.[47].

Most random lasers operate in the highly multimode regime. Individual lasing modes, formed by interference of scattered waves, have distinct frequencies and spatial structures[48,49]. Similar to the case of wave-chaotic cavities, these features result from the non-separable geometry of the system. The spatial field intensity distribution for a lasing mode in a 2D diffusive medium[46] is shown in Fig. 5b. The intensity fluctuates spatially on the wavelength scale, and the mode spreads more or less uniformly over the entire system. When a large number of such modes lase simultaneously with uncorrelated phases, their distinct wave fronts combine incoherently to produce emission with low spatial coherence[50]. The random laser directly generates light of low spatial coherence, in contrast to the conventional approach of reducing spatial coherence of light from a coherent laser by means of, e.g., spinning diffusers. In the latter approach, uncorrelated speckle patterns are *sequentially* formed by a time-varying system (the spinning diffuser) *outside* of the laser cavity, in the random laser, scatterers are directly incorporated *into* the gain medium to *simultaneously* generate many independent speckle patterns, one from each lasing mode.

The number of random lasing modes increases with the scattering strength[27]. As the transport mean free path approaches the emission wavelength, the number of lasing modes rises quickly[45]. Because stronger scattering leads to tighter confinement of random lasing modes, the spatial overlap of the modes is reduced, and their competition for gain becomes weaker. Hence, the spatial coherence of random laser emission can be tuned by varying the density of scatterers, as shown in Fig. 5c-h. Furthermore, the pump beam diameter is typically smaller than the sample size, hence increasing the excitation pump beam size allows more modes to lase[35]. Thus the spatial coherence can also be tuned by changing the pump beam profile[47].

## 2.5. Large-aperture VCSELs

In the preceding sections, the spatial coherence of laser emission was tailored via the cavity geometry or internal structure. These cavity-based approaches are *static*, and aim to manipulate the steady state of lasing. Here we describe a *dynamic* approach that relies on pulsed pumping. The aim is to disrupt the lasing operation so that steady state is not reached. Instead, the laser operates in a transient regime so that spatial coherence cannot build up. This approach can be implemented with large-aperture vertical-cavity surface-emitting lasers (VCSELs). The cavity length is on the order of the emission wavelength, so only one or at most a few longitudinal modes exist[51]. However, there are many more transverse modes, as the diameter of the aperture is much larger than the wavelength. The aperture shape is typically circular, but when it is deformed from the circle, the onset of wave-dynamical chaos is observed[52,53].

When a large-aperture VCSEL is driven by a current pulse, a thermal lens is formed by spatially inhomogeneous heating of the device, leading to an increase of the modal buildup time[54]. In addition, the cavity is constantly expanding and remains non-stationary during the current pulse, causing a strong thermal chirp. Consequently, the cavity modes cannot build up since the cavity environment changes faster than the mode build-up time, and stimulated emission is generated from local islands. The VCSEL operates in a non-modal state, which is a superposition of small independent "coherence islands"[55]. The spatial coherence of emission is greatly reduced, which manifests in the formation of a Gaussian far-field intensity distribution. By tuning the pump pulse duration and amplitude, the degree of spatial coherence can be varied[56]. With a strong thermal lens being established by a constant bias current, a low-amplitude current pulse with 50% duty cycle can create a thermal chirp to break up the global cavity modes. Hence, the spatial coherence of a laser can be drastically changed by temporal modulation of the pump[57].

In addition to spatial coherence control, a current sweep operation scheme was developed to reduce the temporal coherence of VCSELs[58,59]. By taking advantage of thermal effects within the small cavity, application of a rapidly modulated pump current broadens the effective spectrum of a VCSEL, leading to reduced temporal coherence. One advantage of VCSELs for this scheme is that their small gain volume allows rapid thermal changes to produce spectral effects on a short time scale.

## 2.6 Comparison of strategies and implementations

Each of the strategies and implementations for reducing the spatial coherence has its own advantages and problems. The simplest way to reduce the spatial coherence is forming an array of independent lasers, e.g., a 1D array of edge-emitting laser diodes or a 2D array of VCSELs. These lasers must be sufficiently far apart to avoid coupling, thus the total array size and the fabrication cost will rise with the number of lasers. To increase the total number as well as the area density of lasers, it is easier to incorporate all of them in a single cavity, such as a degenerate cavity. An additional advantage of degenerate cavities is that the number of transverse lasing modes can be readily controlled to tune the spatial coherence with little power change. Moreover, a nonlinear crystal can be inserted to the degenerate cavity for intra-cavity frequency doubling[60]. The number of transverse modes at the second harmonic frequency is even larger than that at the fundamental frequency, producing laser emission of low spatial coherence at both frequencies. However, the degenerate cavity laser is relatively bulky and requires careful alignment. A multimode fiber (MMF) amplifier is more robust, compact and has high efficiency, and as an ASE source it exhibits both low spatial and low temporal coherence. However, tuning its coherence is relatively difficult.

While they are very different from conventional lasers, the wave-chaotic microcavity lasers and random lasers are similar in many ways. Both systems have a non-separable geometry due to an asymmetric cavity shape or random refractive index variations, respectively. Most wave-chaotic cavities have 2D geometries and are relatively easy to produce with standard micro-fabrication techniques. Random lasers are more flexible and have been realized in 1D, 2D and 3D geometries. Both random and wave-chaotic lasers are compact and robust, and the number of lasing modes can be changed by spatial modulation of the pump[61-63], enabling spatial coherence control[47]. However, it remains a challenge to have directional emission combined with low spatial coherence from these lasers.

Distinct from the strategies discussed in the preceding paragraphs, temporal modulation of the pump can also greatly reduce the spatial coherence of laser emission, providing a simple method of generating partially coherent, directional beams from large-aperture VCSELs. However, during parts of a pump modulation period there is either no lasing emission or emission with high spatial coherence. The degree of spatial coherence can be periodically modulated at a repetition rate as high[56] as 500 kHz. In conclusion, it strongly depends on a given application which of the laser systems discussed above provides the most suitable low spatial-coherence light source.

## 3. Applications of complex lasers

Many high-speed imaging applications require short detection times, however, it is difficult to achieve low speckle contrast with a short exposure time of the detector. For example, consider a laser with $N$ spatial modes which are frequency degenerate. Each mode has a linewidth of $\delta\omega$, corresponding to a coherence time $\tau_c = 1/\delta\omega$. When illuminating a static scattering medium, each mode generates a distinct speckle pattern. If the exposure time is less than the coherence time, all modes are phase coherent with each other so their scattered fields will add coherently to form a new speckle pattern with unity contrast. However, if the exposure time is much longer than the coherence time, the modes are mutually incoherent and their speckle patterns add in intensity, lowering the contrast to $1/\sqrt{N}$.

If the spatial modes are non-degenerate, their frequency difference will accelerate the loss of coherence. Assuming the frequency spacing $\Delta\omega$ between spatial modes exceeds their linewidth $\delta\omega$, they will be mutually incoherent for an exposure time longer than $1/\Delta\omega$. Therefore, the spectral repulsion of lasing modes in a random medium[32] or a wave-chaotic cavity[20] shortens the exposure time that is needed for speckle suppression.

In a degenerate cavity[12], the frequency spacing between the longitudinal modes $\Delta\omega_l$ is usually much larger than that between the transverse modes $\Delta\omega_t$. On the short time scale of $1/\Delta\omega_l$, all transverse modes with identical longitudinal mode index remain phase coherent, and they interfere to generate a speckle pattern of unity contrast. Each longitudinal mode group produces a distinct speckle pattern, as the phase differences between modes in each group are different. These speckle patterns are mutually incoherent, and an intensity sum of all $M$ independent speckle patterns, where $M$ is the number of longitudinal modes, reduces the speckle contrast to $1/\sqrt{M}$, assuming $M$ is less than the number of transverse modes $N$. On the long time scale of $1/\Delta\omega_t$, all transverse modes can add incoherently to give a speckle contrast of $1/\sqrt{N}$.

Since the spatial coherence of a laser is determined by the number of spatial (transverse) lasing modes, it can be controlled by the cavity geometry and/or internal structure. Generally, the schemes for reducing the spatial coherence, as described in the previous section, are applicable to a variety of lasers that can operate at different wavelengths and with different gain materials. An alternative scheme for lowering the spatial coherence, which is applicable to broadband illumination sources, is to transform temporal incoherence to spatial incoherence via spectral or spatial dispersion[64,65].

In the following, we describe several applications for which lasers with low and/or tunable spatial coherence are needed. Several specific examples with relevant and practical parameters are presented.

3.1. Speckle-free imaging

The ability of a laser with low spatial coherence to suppress speckle noise directly leads to improved image quality, especially when imaging in a scattering environment[66]. Figure 6 shows examples of full-field imaging of an Air Force test chart through a static scattering medium, using four illumination sources. Under spatially coherent illumination with a conventional narrowband laser, interference among scattered photons produces strong speckle noise that corrupts the image beyond recognition. The image is corrupted even when using a broadband laser or an ASE source, which have relatively low temporal coherence. In contrast, when illuminating with a random laser of low spatial coherence, there is little, if any, interference between scattered photons, leading to a uniform background level. Although the scattered photons increase the background level and lower the image contrast, the features of the object remain clearly visible. In a different experiment, it was shown that even a narrow-band random laser, which has high temporal coherence, could provide speckle-free illumination for full-field imaging[67].

In the case of fluorescence imaging, where the fluorescence itself is spatially incoherent and does not produce speckle, the pump light is often a coherent laser beam. The coherent artifacts produced by the pump laser result in spatially non-uniform excitation and cause artificial variations of the fluorescence intensity. To overcome this problem, intense green emission with low spatial coherence is generated via intra-cavity frequency doubling of a solid-state degenerate laser[60]. Using such emission for fluorescence excitation eliminates the coherent artifacts in the full-field fluorescence images.

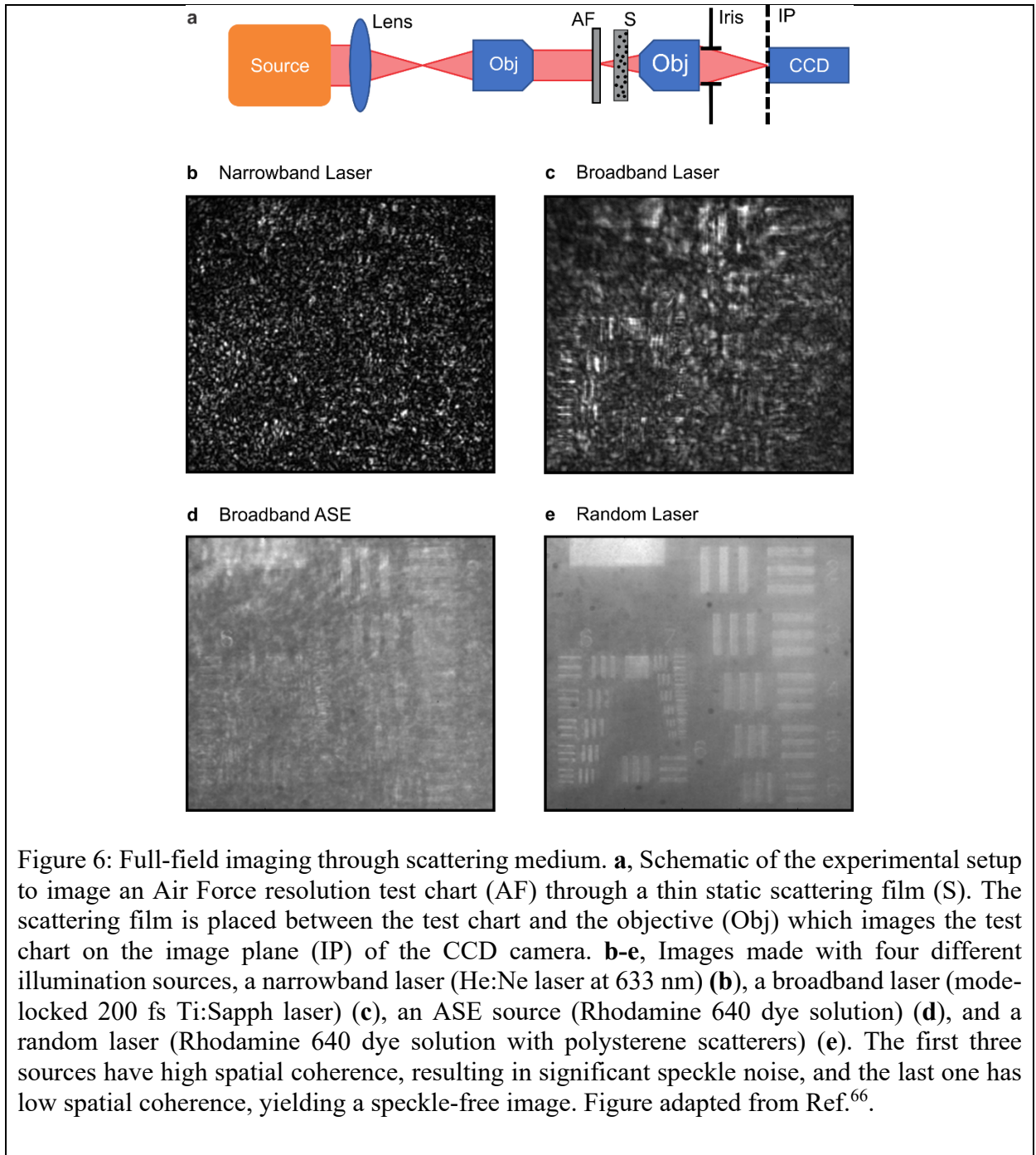

Figure 6: Full-field imaging through scattering medium. **a**, Schematic of the experimental setup to image an Air Force resolution test chart (AF) through a thin static scattering film (S). The scattering film is placed between the test chart and the objective (Obj) which images the test chart on the image plane (IP) of the CCD camera. **b-e**, Images made with four different illumination sources, a narrowband laser (He:Ne laser at 633 nm) **(b)**, a broadband laser (mode-locked 200 fs Ti:Sapph laser) **(c)**, an ASE source (Rhodamine 640 dye solution) **(d)**, and a random laser (Rhodamine 640 dye solution with polysterene scatterers) **(e)**. The first three sources have high spatial coherence, resulting in significant speckle noise, and the last one has low spatial coherence, yielding a speckle-free image. Figure adapted from Ref.[66].

The speckle noise does not only depend on the coherence properties of the illumination source, but also on the imaging optics, including the ratio of the numerical aperture for observation to that for illumination[68]. Accordingly, the required degree of spatial incoherence of an illumination source to suppress speckle noise depends on various parameters of a specific application, such as the amount of scattering, the imaging resolution, etc. For example, a multimode laser could be designed to provide sufficiently low spatial coherence for speckle suppression while maintaining relatively high power per mode as compared to existing spatially incoherent sources such as lamps and LEDs. Typically, about 1000 spatial modes are sufficient to suppress speckles below the level observable by humans[10,11]. Yet lamps and LEDs emit photons into far more modes, thus having lower brightness.

A quantitative measure of the source brightness is the photon degeneracy parameter $\delta$, which gives the number of photons per coherence volume[3]. It is proportional to the spectral radiance, a radiometric measure of the amount of radiation through a unit area and into a unit solid angle within a unit frequency bandwidth. The values of $\delta$ for random lasers and chaotic microcavity lasers are several orders of magnitude higher than those of lamps and LEDs[24]. The greatly improved photon degeneracy allows much shorter exposure times and much higher speed for full-field imaging of transient processes. For example, a random laser can be triggered to produce a short illumination flash at a well-defined delay time, providing uniform, speckle-free background illumination[69]. It enables time-resolved microscopy with an exposure time as short as 10 ns.

3.2. Spatial coherence gating

In addition to speckle-free imaging, a highly multimode laser can provide spatial coherence gating for interferometric detection, enabling parallel confocal image acquisition. In general, confocal microscopy combines high spatial resolution with improved contrast and optical sectioning[70], and traditionally it relies on raster scanning, which limits image acquisition speed[71]. The most common approach to parallelization is to resort to an array of spatially separated pinholes[72], which must be sufficiently separated to prevent cross talk. An alternative approach to completely parallelize confocal image acquisition is to combine interferometric detection with spatial coherence gating[73]. In this approach, each spatial mode (defined by a spatial coherence area) acts as a virtual pinhole. Unlike physical pinholes, these virtual pinholes do not require physical separation to avoid cross talk, enabling parallel acquisition of an entire *en face* plane in a single snapshot without scanning. Although such a microscope is not suitable for fluorescence imaging, it has the potential for high-speed, large-area reflectance imaging with confocal resolution and sectioning. However, it requires a light source that not only has low spatial coherence but also has sufficient power per mode, specifically, a laser with low spatial coherence.

A further advantage of using a laser source with low spatial coherence is that the number of virtual pinholes can be much larger than the number of simultaneous but independent spatial lasing modes. Figure 7 shows an interferometric confocal microscope using an array of 1200 VCSELs coupled to a multimode fiber[74]. The interferometric detection, achieved with an off-axis holography technique, enables parallel acquisition of image information from 18,000 continuous virtual pinholes. The number of virtual pinholes is not limited by the number of emitters in the VCSEL array, as these lasers are combined to eliminate cross talk through averaging, rather than serving as independent imaging channels. Instead, the number of virtual pinholes is determined by the number of resolvable elements provided by the microscope objective, which is determined by the field of view and the point spread function. The interferometric detection of low spatial coherence fields yields a coherent signal that is inherently confocal in the transverse plane and along the axial dimension, thus enhancing both transverse and axial resolution. The microscope in Fig. 7 provides *en face* images with a 210 μm × 280 μm field of view, ~2 μm transverse resolution, and ~8 μm axial resolution in a single shot. The high photon degeneracy shortens the integration times to 100 μs, and increases the 2D frame rate to above 1 kHz. The interferometric detection also recovers the phase of the optical field, which can be used to estimate the height of sub-resolution axial features[75]. The ultrahigh-speed, full-field holographic confocal microscopy is used for *in vivo* quantitative studies of microscale physiology[76].

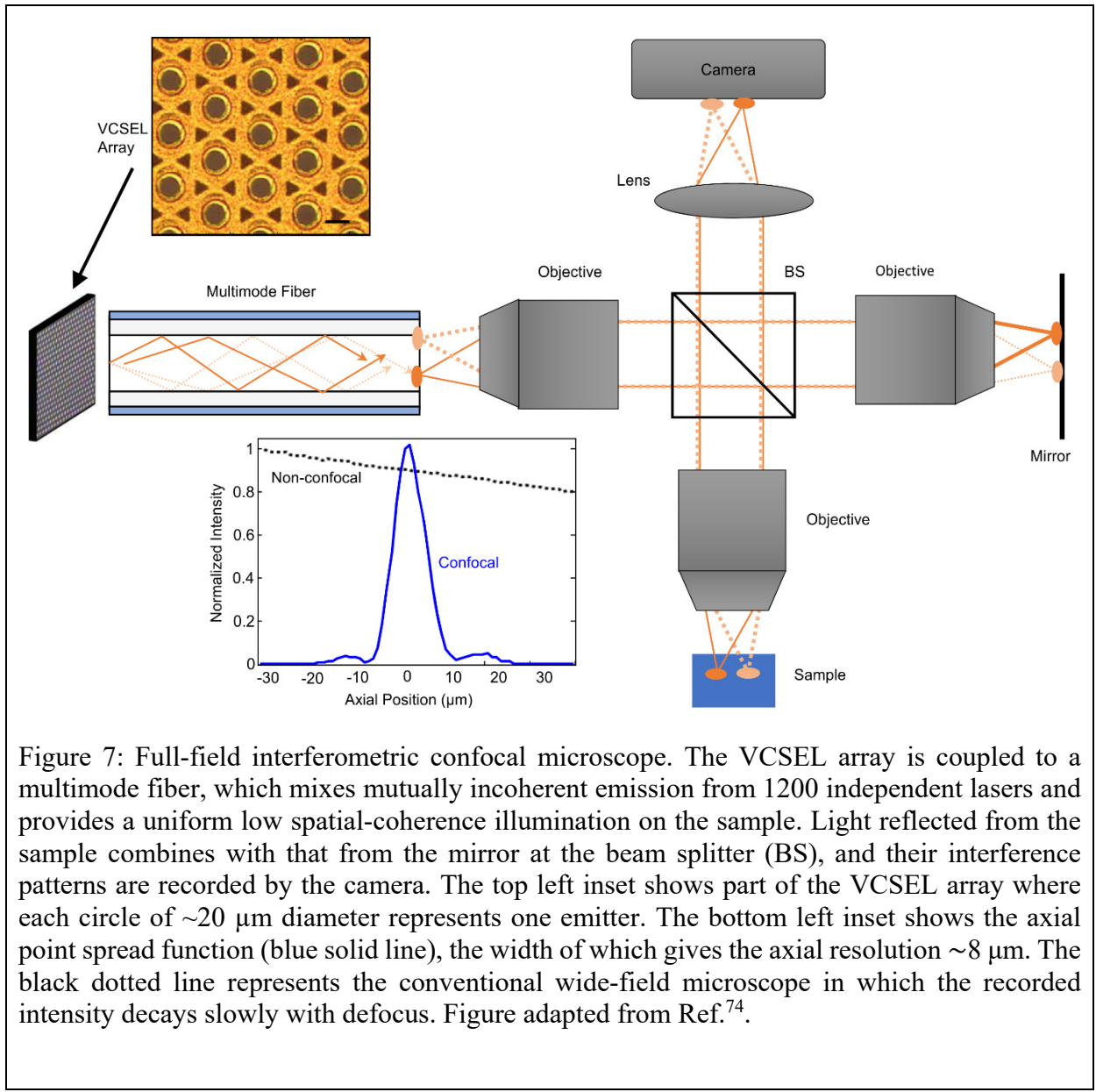

Figure 7: Full-field interferometric confocal microscope. The VCSEL array is coupled to a multimode fiber, which mixes mutually incoherent emission from 1200 independent lasers and provides a uniform low spatial-coherence illumination on the sample. Light reflected from the sample combines with that from the mirror at the beam splitter (BS), and their interference patterns are recorded by the camera. The top left inset shows part of the VCSEL array where each circle of ~20 μm diameter represents one emitter. The bottom left inset shows the axial point spread function (blue solid line), the width of which gives the axial resolution ∼8 μm. The black dotted line represents the conventional wide-field microscope in which the recorded intensity decays slowly with defocus. Figure adapted from Ref.[74].

3.3. Bi-modal imaging

While a speckle-free wide-field image provides the structural information of an object, the speckle formed by random scattering of coherent light carries additional information about the motion of the object. For example, dynamic scatterers in a biological sample, such as moving blood cells, induce time-dependent phase shifts in the scattered light, causing temporal changes in the speckle pattern. Such changes can be used to map the blood flow in living tissues by a technique called laser speckle contrast imaging[77].

Anatomical and functional information on living tissue can be acquired by bimodal imaging, e.g., combining laser speckle contrast microscopy for mapping neural blood flow and an intrinsic signal optical imaging for monitoring tissue oxygenation. This was achieved by using two separate illumination sources, a laser and a LED, and switching rapidly between them[58]. However, a laser and a LED typically have very different beam divergence angles and power densities, and separate optical setups are necessary for beam collimation and power adjustment to ensure that the illumination area and the light intensity do not change much when switching the two sources. Therefore, a single illumination source with tunable spatial coherence but constant power and beam size would be advantageous for such applications. It could be possible to increase the spatial coherence of a LED by spatial filtering, but then the power would be greatly reduced. A better, and probably the best, scenario is to tune the spatial coherence of a laser, such as a degenerate cavity laser, with little power loss[16,17].

An electrically-pumped semiconductor-based degenerate VECSEL (vertical external cavity surface emitting laser) with continuous wave (CW) emission was developed for imaging the embryo heart function in a Xenopus, an important animal model of heart disease[18]. The spatial coherence of the laser emission was switched from low to high, so that the low-spatial-coherence illumination was used for high-speed (100 frames per second) speckle-free imaging of the dynamic heart structure, and the high spatial-coherence illumination for laser speckle contrast imaging of the blood flow.

3.4. Laser wavefront shaping

A unique feature of the degenerate cavity laser is that it can produce arbitrary output beam profiles. Beam shaping is a key technology for the applications of high power lasers to materials processing and device fabrications[78]. External shaping of a laser beam (outside of the laser cavity) to a desired transverse profile was implemented with a variety of elements and techniques including diffractive optical elements, free-form optics, and digital holography with spatial light modulators. For example, an aspheric or diffractive element can transform a Gaussian intensity profile of a single-mode laser beam to a flat-top beam.

A multimode degenerate cavity laser can directly generate an output beam with flat-top profile[60]. This is due to the unique feature of the degenerate cavity laser that it can support a large number of transverse modes with almost identical frequencies but different spatial wave fronts. Using intra-cavity elements, it is possible to superimpose these modes to produce arbitrary output beam profiles.

Due to diffraction, the profiles of light beams vary as they propagate. By manipulating the spatial coherence, such variations of the beam profiles can be controlled[79]. In the 4f degenerate cavity laser (Fig. 2a), direct access to both the real space and the Fourier space of the lasing modes enables simultaneous control of the beam profile and the spatial coherence. Shaped beams with tailored spatial coherence can be obtained by inserting an amplitude mask near one of the mirrors and a Fourier mask in the mutual focal plane of the two lenses[80]. If the Fourier mask is a circular aperture, decreasing the aperture diameter blocks the large transverse wave vector components, so the output beam experiences less diffraction upon propagation but the minimal feature size of its spatial profile is increased. It is possible to produce a shaped beam that is propagation invariant by resorting to an annular aperture Fourier mask (see Fig. 8a) to obtain a Bessel spatial coherence function of the emission. The output beam, regardless of its transverse profile (which is set by an intra-cavity amplitude mask), can propagate over a relatively long distance with minimal diff-raction, as shown in Fig. 8b. Geometric phase metasurfaces can also be incorporated into a degenerate cavity laser as an output coupler to efficiently generate spin-dependent twisted light beams of different topologies[81].

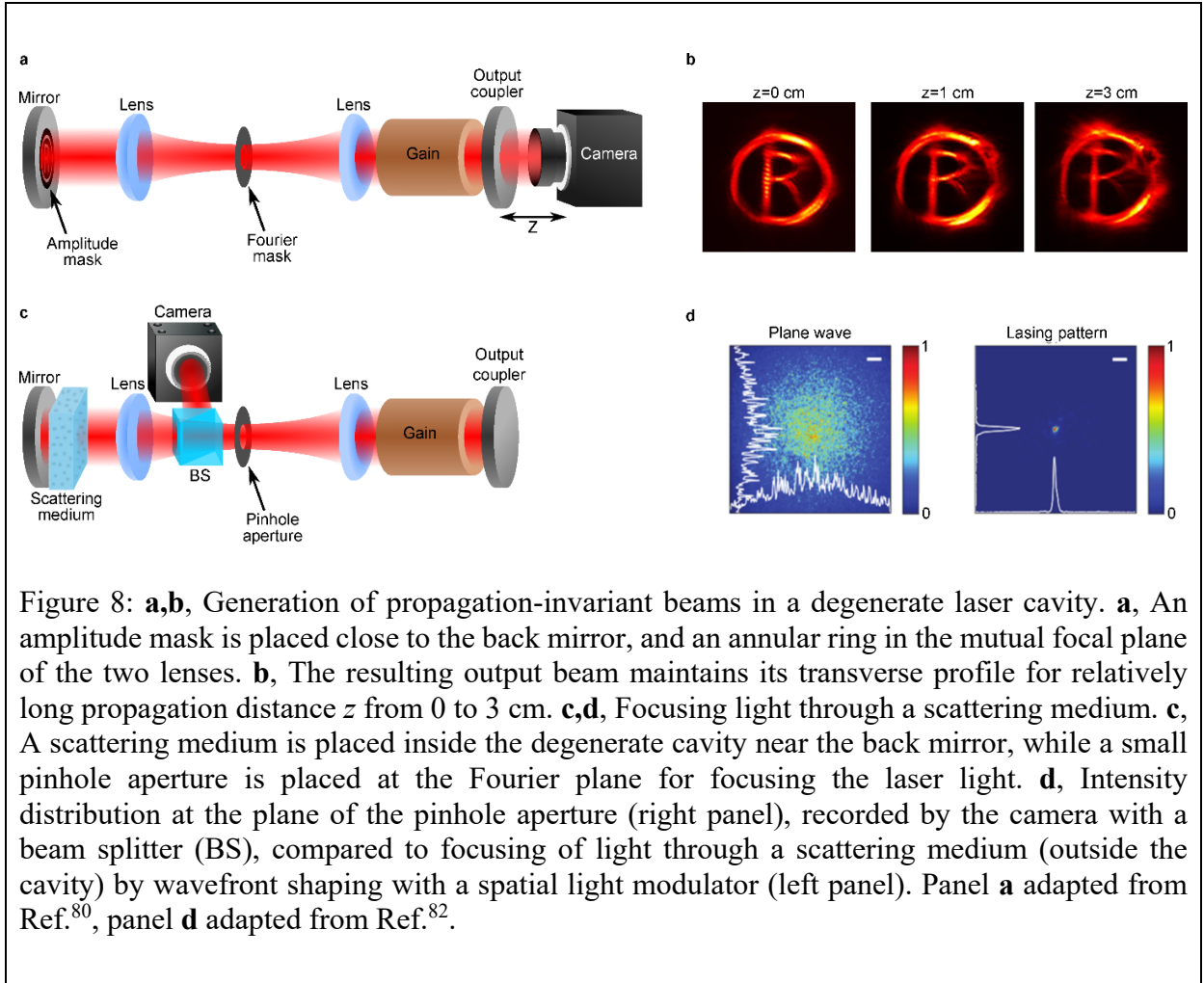

Figure 8: **a,b**, Generation of propagation-invariant beams in a degenerate laser cavity. **a**, An amplitude mask is placed close to the back mirror, and an annular ring in the mutual focal plane of the two lenses. **b**, The resulting output beam maintains its transverse profile for relatively long propagation distance *z* from 0 to 3 cm. **c,d**, Focusing light through a scattering medium. **c**, A scattering medium is placed inside the degenerate cavity near the back mirror, while a small pinhole aperture is placed at the Fourier plane for focusing the laser light. **d**, Intensity distribution at the plane of the pinhole aperture (right panel), recorded by the camera with a beam splitter (BS), compared to focusing of light through a scattering medium (outside the cavity) by wavefront shaping with a spatial light modulator (left panel). Panel **a** adapted from Ref.[80], panel **d** adapted from Ref.[82].

In addition to shaping the output beam, the intra-cavity wave front can be tailored for efficiently propagating through a random scattering medium, which is placed inside a degenerate cavity laser[82]. The light scattered by a rapidly varying disordered sample was focused at sub-microsecond timescales, without requiring a computer-controlled spatial light modulator (SLM) or electronic feedback. This was implemented by inserting a pinhole in the mutual focal plane of the two lenses in the degenerate cavity, and the lasing process found the wave front that focused the maximum power onto the pinhole in order to minimize the loss (see Fig. 8c,d). This approach relies on the self-organization of the optical field inside the degenerate cavity to create the optimal wave front that forms a sharp focus from the otherwise randomly scattered light. This wave front effectively compensates the effect of scattering induced by the random medium. Such wavefront shaping is achieved by all-optical feedback, and is therefore orders of magnitude faster than all other wavefront shaping techniques.

## 4. Conclusions and outlook

Unlike the relatively simple configurations of traditional laser resonators, the configurations and geometries of the unconventional lasers covered in this review are complex. Despite their differences, all these complex lasers possess a large number of spatial degrees of freedom, allowing for novel properties and characteristics. By tailoring the spatial structures of lasing modes, their nonlinear interactions via the gain material can be controlled. Consequently, the number of lasing modes and the spatial coherence of emission can be tuned over a wide range. Moreover, the output beams can have arbitrary profile and topology. The different types of complex lasers reviewed here each have unique properties and are suitable for various applications with specific requirements, including high-speed speckle-free imaging, spatial coherence gating and focusing through scattering media.

The complex lasers have additional advantages over traditional lasers. The wave-chaotic microcavity lasers, as well as the random lasers, can suppress spatio-temporal instabilities and chaotic dynamics that are common for high-power lasers[83]. Wave interference effects in these lasers disrupt nonlinear processes that form self-organized structures such as filaments, which are inherently unstable, thereby resulting in stable lasing dynamics. In a multimode fiber laser, spatio-temporal mode-locking is achieved[84], paving the way for full control of the spatio-temporal coherence.

The "passive" schemes for manipulating laser performance via the cavity geometry and internal structure could be combined with "active" control with gain and/or loss[85]. Adaptive shaping of the spatial distribution of the pump intensity enables not only single-mode lasing at any selected wavelength[62,86], but also switching of emission directions[87,88]. A symmetric arrangement of gain and loss in a microcavity (Parity-Time symmetry) results in stable single-mode operation[89,90]. Optical loss can also induce suppression and revival of lasing in the vicinity of an exceptional point[91,92].

Since complex lasers have numerous degrees of freedom, they may be used for reservoir computing[93]. The self-adaptive nature of a highly multimode laser is exploited for rapid phase retrieval[94]. Imposing constraints in a digital degenerate cavity laser breaks the degeneracy between the transverse modes and forces the system to find a lasing state with minimal loss. In this manner, complex lasers can be mapped to hard computational problems and used as physical simulators[95].

While this review focuses on the first order coherence of complex lasers, the second order or even higher order coherence properties are also interesting to explore in the future. In a laser, optical gain saturation suppresses intensity fluctuations of the emission, enhancing the second-order coherence. The nonlinear multimodal interactions can make the higher-order coherence properties of a complex laser very different from those of a conventional laser.

To conclude, the study of complex lasers bridges multiple disciplines, including mesoscopic physics, nonlinear dynamics, quantum optics, wave-dynamical chaos, and non-Hermitian physics. Thanks to their diversity and versatility, such lasers constitute a toolbox for various applications, allowing application-driven laser design. This article reviews certain processes and applications of complex laser systems, however, it only hints at the enormous and largely unexplored potential of these systems.

**Acknowledgements**

H.C., R.C., A.F. and N.D. acknowledge funding by the United States – Israel Binational Science Foundation (BSF) under grant no. 2015509. R.C., A.F. and N.D. were supported by the Israel Science Foundation (ISF) by grant no. 1881/17. H.C. and S.B. acknowledge support by the Office of Naval Research (ONR) via grant no. N00014-13-1-0649. H.C. was supported by the Air Force Office of Scientific Research under grant no. FA9550-16-1-0416.